\documentclass[prd,aps]{revtex4}
\begin{document}
\title{Angular momentum content of a virtual graviton}
\author{Duane A. Dicus}
\affiliation{Center for Particle Physics, University of Texas at Austin, Austin, TX 78712}
\author{Scott Willenbrock}
\affiliation{Department of Physics, University of Illinois at Urbana-Champaign,
1110 West Green Street, Urbana, IL 61801}

\date{\today}

\begin{abstract}
We show that a virtual graviton has a $J=0$ component, which
serves to cancel the $J=2,J_z=0$ component when the graviton is on
shell.  In contrast, a massive graviton has no $J=0$ component
either on or off shell.  This difference is responsible for the
van Dam-Veltman-Zakharov discontinuity.
\end{abstract}

\maketitle

\section{Introduction}

It was recently noticed, in the consideration of $2\to 2$
scattering amplitudes via graviton exchange, that a virtual
graviton has a component with zero angular momentum ($J=0$)
\cite{Datta:2003kn,Datta:2004mr,Han:2004wt}. While this is
forbidden for a real graviton, which has $J=2$, $J_z=\pm 2$, a
$J=0$ component of a virtual graviton does not violate any
fundamental principle. Nevertheless, it is perhaps surprising,
given that a virtual photon does not have a $J=0$ component.
Furthermore, a {\em massive} graviton does not have a $J=0$
component either on or off shell \cite{Datta:2003kn}, so it seems
odd that a virtual massless graviton has a $J=0$ component.

The fact that the graviton propagator has a $J=0$ component was
already shown in Ref.~\cite{VanNieuwenhuizen:1973fi}.  In this
note we decompose the graviton propagator in terms of polarization
tensors and explicitly display the $J=0$ component.  We then show
that no such component is present in the massive graviton
propagator. This is the reason for the van Dam-Veltman-Zakharov
discontinuity between the massive and massless graviton
propagators \cite{vanDam:1970vg,Zakharov}.

We begin in Section~\ref{sec:photon} with a brief reminder about
virtual photons.  Virtual gravitons are dealt with in
Section~\ref{sec:graviton}, followed by massive gravitons in
Section~\ref{sec:massive}.  We conclude with a discussion of a
recent claim that angular momentum conservation is violated in
quantum gravity \cite{Datta:2003kn,Datta:2004mr}.

\section{Virtual photons}\label{sec:photon}

We begin by considering a virtual photon of four-momentum
$q^\mu=(\omega,0,0,\kappa)$.  The photon propagator in $R_\xi$
gauge is
\begin{equation}
D^{\mu\nu}=\left(-g^{\mu\nu}+(1-\xi)\frac{q^\mu
q^\nu}{q^2}\right)\frac{i}{q^2+i\epsilon}\;.
\end{equation}
The propagator connects two currents, both of which are conserved.
The numerator of the amplitude is proportional to
\begin{equation}
J_{A\mu}D^{\mu\nu}J_{B\nu} \sim
-J^{0}_AJ^{0}_B+J^{1}_AJ^{1}_B+J^{2}_AJ^{2}_B+J^{3}_AJ^{3}_B
\label{eq:vector}
\end{equation}
where we have used current conservation, $q^\mu J_\mu =0$, to
eliminate the gauge-dependent part of the propagator.
Alternatively, one could simply work in `t Hooft-Feynman gauge
($\xi=1$).

Current conservation implies
\begin{equation}
\omega J^0=\kappa J^3 \label{eq:current}
\end{equation}
for both $J_A$ and $J_B$.  For a real photon ($\omega=\kappa$),
this gives $J^0=J^3$, and Eq.~(\ref{eq:vector}) reduces to
\begin{equation}
J_{A\mu}D^{\mu\nu}J_{B\nu} \sim J^{1}_AJ^{1}_B+J^{2}_AJ^{2}_B
\end{equation}
which shows that a real photon has only transverse polarizations.
A virtual photon, however, also has a longitudinal component. The
transverse and longitudinal polarization vectors are
\begin{eqnarray*}
\epsilon^{\mu}_{+}&=&\frac{1}{\sqrt 2}(0,-1,-i,0) \\
\epsilon^{\mu}_{-}&=&\frac{1}{\sqrt 2}(0,1,-i,0)\\
\epsilon^{\mu}_{0}&=&\frac{1}{\sqrt{q^2}}(\kappa,0,0,\omega)
\end{eqnarray*}
which correspond to $J=1$, $J_z=+1,-1,0$, respectively. Through
the use of current conservation, Eq.~(\ref{eq:current}),
Eq.~(\ref{eq:vector}) may be written in terms of these
polarization vectors as
\begin{equation}
J_{A\mu}D^{\mu\nu}J_{B\nu} \sim
J_{A\mu}\sum_{\lambda=-1}^{1}\epsilon^{\mu}_{\lambda}\epsilon^{\nu
*}_{\lambda}J_{B\nu} \label{eq:vector2}
\end{equation}
where the sum runs over the three $J=1$ polarization vectors given
above \footnote{In this expression, one should treat $\sqrt{q^2}$
formally as a real number, even when the photon is spacelike
($q^2<0$).}. In particular, there is no $J=0$ component.

\section{Virtual gravitons}\label{sec:graviton}

We now perform a similar analysis for the graviton propagator
\cite{Veltman:vx,Feynman:kb},
\begin{equation}
D^{\mu\nu\rho\sigma}=\frac{1}{2}\left(g^{\mu\rho}g^{\nu\sigma}+g^{\mu\sigma}g^{\nu\rho}
-g^{\mu\nu}g^{\rho\sigma}\right)\frac{i}{q^2+i\epsilon}\;.
\label{eq:prop}
\end{equation}
where we have specialized to harmonic (De Donder) gauge.  The
graviton propagator connects two conserved energy-momentum
tensors,
\begin{eqnarray}
T_{A\mu\nu}D^{\mu\nu\rho\sigma}T_{B\rho\sigma}& \sim
&\frac{1}{2}T^{00}_A[T^{00}_B
+T^{11}_B+T^{22}_B+T^{33}_B] \nonumber \\
   & &+\frac{1}{2}T^{11}_A[T^{00}_B+T^{11}_B-T^{22}_B-T^{33}_B] \nonumber \\
   & &+\frac{1}{2}T^{22}_A[T^{00}_B-T^{11}_B+T^{22}_B-T^{33}_B] \nonumber \\
   & &+\frac{1}{2}T^{33}_A[T^{00}_B-T^{11}_B-T^{22}_B+T^{33}_B] \nonumber
   \\
   & &-2T^{01}_AT^{01}_B-2T^{02}_AT^{02}_B-2T^{03}_AT^{03}_B \nonumber
   \\
   & &+2T^{12}_AT^{12}_B+2T^{13}_AT^{13}_B+2T^{23}_AT^{23}_B\;,\label{eq:tensor}
\end{eqnarray}
where we display only the numerator on the right-hand side. The
symmetry $T^{\mu\nu}=T^{\nu\mu}$ has been used to obtain this
expression.  In a more general gauge, terms proportional to the
graviton four-momentum may be neglected due to conservation of the
energy-momentum tensor, $q^\mu T_{\mu\nu}=0$.

Conservation of the energy-momentum tensor implies
\begin{equation}
\omega T^{0\nu}=\kappa T^{3\nu}\;. \label{eq:cons}
\end{equation}
For a real graviton ($\omega=\kappa$), this gives
$T^{0\nu}=T^{3\nu}$, and Eq.~(\ref{eq:tensor}) reduces to
\begin{equation}
T_{A\mu\nu}D^{\mu\nu\rho\sigma}T_{B\rho\sigma} \sim
\frac{1}{2}[(T^{11}_A-T^{22}_A)(T^{11}_B-T^{22}_B)]+2T^{12}_AT^{12}_B
\label{eq:real}
\end{equation} which shows that a real graviton has
only transverse components. A virtual graviton, however, has
additional components.  To identify them, we first rewrite
Eq.~(\ref{eq:tensor}) as
\begin{eqnarray}
T_{A\mu\nu}D^{\mu\nu\rho\sigma}T_{B\rho\sigma}& \sim
&\frac{1}{2}[(T^{11}_A-T^{22}_A)(T^{11}_B-T^{22}_B)]+2T^{12}_AT^{12}_B \nonumber \\
   & &+2[T^{13}_AT^{13}_B+T^{23}_AT^{23}_B-T^{01}_AT^{01}_B-T^{02}_AT^{02}_B] \nonumber \\
   & &+\frac{1}{6}[2(T^{00}_A-T^{33}_A)+T^{11}_A+T^{22}_A][2(T^{00}_B-T^{33}_B)+T^{11}_B+T^{22}_B] \nonumber \\
   & &-\frac{1}{6}[-T^{00}_A+T^{33}_A+T^{11}_A+T^{22}_A][-T^{00}_B+T^{33}_B+T^{11}_B+T^{22}_B]
\label{eq:tensor2}
\end{eqnarray}
where current conservation, Eq.~(\ref{eq:cons}), has been used. We
now proceed to identify each line above with a particular
polarization.

An easy way to construct $J=2$ polarization tensors is to take
products of the $J=1$ polarization vectors given in the previous
section.  The $J=2$, $J_z=\pm 2$ polarization tensors are given by
\begin{equation}
\epsilon^{\mu\nu}_{2,\pm2}=\epsilon^{\mu}_{\pm}\epsilon^{\nu}_{\pm}=
\frac{1}{2}\left(\begin{array}{cccc}
                                    0 & 0 & 0 & 0  \\
                                    0 & 1 & \pm\,i & 0  \\
                                    0 & \pm\,i & -1 & 0  \\
                                    0 & 0 & 0 & 0   \\  \end{array}\right)\;.
\label{eq:J22}
\end{equation}
Using raising and lowering operators or, equivalently,
Clebsch-Gordan coefficents \cite{Eidelman:2004wy}, we derive the
$J=2$, $J_z=\pm 1,0$ polarization tensors
\begin{equation}
\epsilon^{\mu\nu}_{2,\pm1}=\frac{1}{\sqrt
2}(\epsilon^{\mu}_{\pm}\epsilon^{\nu}_{3}+\epsilon^{\mu}_{3}\epsilon^{\nu}_{\pm})
=\frac{1}{2\sqrt{q^2}}\left(\begin{array}{cccc}
                                                   0 & \mp\kappa & -i\kappa & 0  \\
                                                  \mp\kappa & 0 & 0 & \mp\omega  \\
                                                 -i\kappa & 0 & 0 & -i\omega  \\
                                                   0 & \mp\omega & -i\omega & 0  \\   \end{array}\right)\,.
\label{eq:J21}
\end{equation}
\begin{equation}
\epsilon^{\mu\nu}_{2,0}=\frac{1}{\sqrt
6}(\epsilon^{\mu}_{+}\epsilon^{\nu}_{-}+\epsilon^{\mu}_{-}\epsilon^{\nu}_{+}
+2\epsilon^{\mu}_{3}\epsilon^{\nu}_{3}) =\frac{1}{\sqrt
6}\left(\begin{array}{cccc}
                                               2\frac{\kappa^2}{q^2} & 0 & 0 & 2\frac{\kappa\omega}{q^2}  \\
                                                         0           & -1 & 0 & 0       \\
                                                         0           & 0 & -1 & 0       \\
                                               2\frac{\kappa\omega}{q^2} & 0 & 0 & 2\frac{\omega^2}{q^2}  \\  \end{array}\right)\,,
\label{eq:J20}
\end{equation}
and also the $J=0$ polarization tensor
\begin{equation}
\epsilon^{\mu\nu}_{0,0}=\frac{1}{\sqrt
3}(\epsilon^{\mu}_{+}\epsilon^{\nu}_{-}+\epsilon^{\mu}_{-}\epsilon^{\nu}_{+}
-\epsilon^{\mu}_{3}\epsilon^{\nu}_{3}) =\frac{1}{\sqrt
3}\left(\begin{array}{cccc}
                                               -\frac{\kappa^2}{q^2} & 0 & 0 & -\frac{\kappa\omega}{q^2}  \\
                                                          0          & -1 & 0 & 0    \\
                                                          0          & 0 & -1 & 0  \\
                                               -\frac{\kappa\omega}{q^2} & 0 & 0 & -\frac{\omega^2}{q^2}  \\  \end{array}\right)\,.
\label{eq:J00}
\end{equation}
Using these polarization tensors, we can write
Eq.~(\ref{eq:tensor2}) as \footnote{See the previous footnote.}
\begin{equation}
T_{A\mu\nu}D^{\mu\nu\rho\sigma}T_{B\rho\sigma} \sim
T_{A\mu\nu}\left(\sum_{\lambda=-2}^{2}\epsilon^{\mu\nu}_{2,\lambda}\epsilon^{\rho\sigma*}_{2,\lambda}
-\frac{1}{2}\epsilon^{\mu\nu}_{0,0}\epsilon^{\rho\sigma*}_{0,0}\right)T_{B\rho\sigma}\;.
\label{eq:result}
\end{equation}
The first line of Eq.~(\ref{eq:tensor2}) [which is identical to
Eq.~(\ref{eq:real})] corresponds to the $J=2$, $J_z=\pm 2$
contribution, which is present both on and off shell. Current
conservation, Eq.~(\ref{eq:cons}), must be used to make this
correspondence, as well as the correspondences below. The second
line corresponds to the $J=2$, $J_z=\pm 1$ contribution. It is
evident that this vanishes on shell, where $T^{0\nu}=T^{3\nu}$.
The third and fourth lines correspond to the $J=2, J_z=0$ and
$J=0$ contributions, respectively. Although neither of these
vanish on shell, it is evident that they cancel on shell
($T^{00}=T^{33}$). Thus the $J=0$ component is necessary in order
to ensure that a real graviton has no $J=2$, $J_z=0$ component. It
is evident from Eq.~(\ref{eq:tensor2}) that the $J=0$ component
couples to the trace of the energy-momentum tensor.

The coefficient of the $J=0$ contribution to Eq.~(\ref{eq:result})
is unconventional in both its magnitude and its sign, but since
this contribution does not contribute on shell, it does not
violate any fundamental principles.  In particular, the negative
sign does not indicate a ghost, because there is no pole
associated with the $J=0$ contribution.  Thus we conclude that a
$J=0$ component of the graviton propagator is present, and that it
serves the purpose of canceling the $J=2, J_z=0$ contribution on
shell \cite{VanNieuwenhuizen:1973fi}.

\section{Massive gravitons}\label{sec:massive}

The propagator of a massive graviton of mass $M$ is
\cite{Veltman:vx}
\begin{equation}
D_{M}^{\mu\nu\rho\sigma}=\frac{1}{2}\left(\bar g^{\mu\rho}\bar
g^{\nu\sigma}+\bar g^{\mu\sigma}\bar g^{\nu\rho}-\frac{2}{3}\bar
g^{\mu\nu}\bar g^{\rho\sigma}\right)\frac{i}{q^2-M^2+i\epsilon}
\end{equation}
where
\begin{equation}
\bar g^{\mu\nu}=g^{\mu\nu}-\frac{q^\mu q^\nu}{M^2}\;.
\end{equation}
When sandwiched between two conserved energy-momentum tensors,
terms proportional to the graviton four-momentum vanish, and $\bar
g^{\mu\nu}$ may be replaced by $g^{\mu\nu}$.  The numerator of the
massive graviton propagator is then identical to that of
Eq.~(\ref{eq:prop}) for a massless graviton, with the exception of
the coefficient of the last term. This gives rise to the van
Dam-Veltman-Zakharov discontinuity
\cite{vanDam:1970vg,Zakharov,Vainshtein:1972sx,Deffayet:2001uk,Arkani-Hamed:2002sp}.

The numerator of the massive graviton propagator may be written in
terms of massive $J=2$ polarization tensors as
\begin{equation}
D_M^{\mu\nu\rho\sigma}\sim
\sum_{\lambda=-2}^{2}\epsilon^{\mu\nu}_{2,\lambda}\epsilon^{\rho\sigma*}_{2,\lambda}
\end{equation}
where the massive polarization tensors are identical to those of
the massless graviton, Eqs.~(\ref{eq:J22}) -- (\ref{eq:J20}), but
with $q^2$ replaced by $M^2$ throughout.  We see that, unlike the
massless graviton propagator, Eq.~(\ref{eq:result}), there is no
$J=0$ contribution. In the massless case, the $J=0$ contribution
was necessary to cancel the $J=2$, $J_z=0$ contribution on shell.
In the massive case no such cancellation is necessary, as an
on-shell massive graviton has a $J=2$, $J_z=0$ component.

\section{Discussion}

Recently a pair of papers have appeared that argue that the
presence of a $J=0$ contribution in the graviton propagator
implies a violation of angular momentum conservation
\cite{Datta:2003kn,Datta:2004mr}. In the latter paper, it is
argued that although a partial-wave analysis reveals a $J=0$
component in certain $2\to 2$ amplitudes, an explicit
decomposition of that amplitude in terms of polarization tensors
shows that only $J=2$, $J_3=\pm 2$ components are present
\footnote{Actually, $J=2$, $J_3=0$ and $J=0$ components are also
present, but happen to cancel for these particular amplitudes.}.
However, while the partial-wave analysis refers to the angular
momentum along the collision axis, $z$, the decomposition is
performed in terms of polarization tensors along an axis
orthogonal to the scattering plane ($J_3\neq J_z$).  We will show
that when the amplitude is decomposed along the $z$ axis, a $J=0$
component is indeed present, consistent with the partial-wave
analysis. Thus there is no contradiction, and hence no evidence
for violation of angular momentum conservation.

The amplitudes studied in Ref.~\cite{Datta:2004mr} involve
same-helicity massive fermion-antifermion or massive vector
bosons.  It is shown that in both cases the energy-momentum tensor
with a virtual graviton, in the center-of-momentum frame, is
proportional to
\begin{equation}
\hat T^{ij}[f_\pm,\bar f_\pm] \sim \hat T^{ij}[V_\pm,V_\pm]\sim
{\bf k}^i{\bf k}^j
\end{equation}
where ${\bf k}$ is the three-momentum of one of the external
particles.  If we decompose the angular momentum along the
collision axis, $z$, one obtains
\begin{equation}
{\bf k}^i{\bf k}^j=\left(\begin{array}{ccc}

                                                         0           & 0 & 0        \\
                                                         0           & 0 & 0        \\
                                                0 & 0 & 1  \\  \end{array}\right)
                                               =\frac{1}{\sqrt
                                               3}(\sqrt 2{\bf
                                               \epsilon}_{2,0}^{ij}-{\bf
                                               \epsilon}_{0,0}^{ij})
\end{equation}
where ${\bf \epsilon}_{J,\lambda}$ are the spatial components of
the polarization tensors of Eqs.~(\ref{eq:J20}) and (\ref{eq:J00})
in the center-of-momentum frame ($\kappa=0$). This agrees with the
partial-wave analysis of Ref.~\cite{Datta:2004mr}, which reveals a
linear combination of $J=2,J_z=0$ and $J=0$ contributions.

In this paper we have shown that a virtual graviton has a $J=0$
component, and that it serves the purpose of canceling the $J=2$,
$J_z=0$ component when the graviton is on shell.  In contrast, a
massive graviton has no $J=0$ component either on or off shell.
This difference gives rise to the van Dam-Veltman-Zakharov
discontinuity.

\section*{Acknowledgements}

We are grateful for conversations with T.~Han and I.~Rothstein.
This work was supported in part by the U.~S.~Department of Energy
under contracts Nos.~DE-FG03-93ER40757 and DE-FG02-91ER40677.

\end{document}